# Equivalence and Hermiticity of Dirac Hamiltonians in the Kerr gravitational field

M.V. Gorbatenko, V.P. Neznamov[1]

RFNC-VNIIEF, 37 Mira Ave., Sarov, 607188, Russia

## Abstract

In the paper, for the Kerr field, we prove that Chandrasekhar's Dirac Hamiltonian and the self-adjoint Hamiltonian $H_\eta$ with a flat scalar product of the wave functions are physically equivalent.

Operators of transformation of Chandrasekhar's Hamiltonian and wave functions to the $\eta$ representation with a flat scalar product are defined explicitly.

If the domain of the wave functions of Dirac's equation in the Kerr field is bounded by two-dimensional surfaces of revolution around the $z$ axis, Chandrasekhar's Hamiltonian and the self-adjoint Hamiltonian in the $\eta$ representation are Hermitian with equality of the scalar products, $(\psi, H\varphi) = (H\psi, \varphi)$.

[1] E-mail: neznamov@vniief.ru

## 1. Introduction

Chandrasekhar in [1] separated angular and radial variables of Dirac's equation in the Kerr gravitational field [2]. Page in [3] extended Chandrasekhar's approach to Dirac's equation in the Kerr-Newman field [4]. In both cases, they used metrics in the Boyer-Lindquist coordinates [5].

Chandrasekhar's and Page's stationary four-component Hamiltonians are pseudo-Hermitian [6] - [8], [11] or, in other words, Hermitian with Parker's weight operator [9] in scalar products of the wave functions.

In [10] - [12], we developed a method for deriving self-adjoint Dirac Hamiltonians with a flat scalar product of the wave functions within the framework of pseudo-Hermitian quantum mechanics [6] - [8] for arbitrary, including time dependent, gravitational fields[2]. In particular, in [12], we obtained a self-adjoint Hamiltonian in the Kerr gravitational field. This Hamiltonian can be easily extended to Kerr-Newman geometry.

In the present paper, we prove that Chandrasekhar's Hamiltonian is equivalent to the corresponding self-adjoint Hamiltonian in the $\eta$ representation. For Kerr geometry, an explicit form of operators of transformation to the $\eta$ representation is defined. We declared the above equivalence in [13] without proof because of cumbersome mathematical espressions.

In this paper also for the domains of the wave functions of Dirac's equation in the Kerr field, bounded by two-dimensional surfaces of revolution around the $z$ axis, we prove the fulfillment of the Hermiticity condition for Chandrasekhar's Hamiltonian and the Hamiltonian in the $\eta$ representation. Such surfaces include, in particular, the surfaces of the outer and the inner ergospheres of the Kerr field.

## 2. Kerr metric in the Boyer-Lindquist coordinates

The Kerr solution to the general relativity equations is characterized by a point source of gravitational field of mass $M$, which rotates with an angular momentum $\mathbf{J} = Mc\mathbf{a}$, where $c$ is the speed of light, $\mathbf{a}$ is the reduced angular momentum. For the Kerr solution $\mathbf{a} = (0, 0, a)$.

Below we use the units in which $\hbar = c = 1$.

[2] Below such representation of Dirac Hamiltonians is called the $\eta$ - representation.

Tetrad vectors $H_{\underline{\alpha}}^{\mu}$ are defined by the relationship

$$H_{\underline{\alpha}}^{\mu} H_{\underline{\beta}}^{\nu} g_{\mu\nu} = \eta_{\underline{\alpha\beta}}, \qquad (1)^3$$

where

$$\eta_{\underline{\alpha\beta}} = diag[1,-1,-1,-1]. \qquad (2)$$

The global (non-underlined) indices are lowered and raised up by means of the metric tensor $g_{\mu\nu}$ and the inverse tensor $g^{\mu\nu}$, and the local (underlined) indices, by means of the tensors $\eta_{\underline{\alpha\beta}}$, $\eta^{\underline{\alpha\beta}}$.

Dirac's matrices satisfy the relationships

$$\gamma^{\mu}\gamma^{\alpha} + \gamma^{\alpha}\gamma^{\mu} = 2g^{\mu\alpha}E, \qquad (3)$$

$$\gamma^{\underline{\mu}}\gamma^{\underline{\alpha}} + \gamma^{\underline{\alpha}}\gamma^{\underline{\mu}} = 2\eta^{\underline{\mu\alpha}}E, \qquad (4)$$

where $E$ is a 4 x 4 identity matrix. The relationship between $\gamma^{\alpha}$ and $\gamma^{\underline{\alpha}}$ is given by the expression

$$\gamma^{\alpha} = H_{\underline{\beta}}^{\alpha}\gamma^{\underline{\beta}}. \qquad (5)$$

The Kerr metric in the Boyer-Lindquist coordinates $(t,r,\theta,\varphi)$ has the form

$$ds^2 = \left(1 - \frac{r_0 r}{\rho_K^2}\right)dt^2 + \frac{2ar_0 r}{\rho_K^2}\sin^2\theta dt d\varphi - \frac{\rho_K^2}{\Delta}dr^2 - \rho_K^2 d\theta^2 - \\ -\left(r^2 + a^2 + \frac{a^2 r_0 r}{\rho_K^2}\sin^2\theta\right)\sin^2\theta d\varphi^2. \qquad (6)$$

In (6), $r_0 = \dfrac{2GM}{c^2}$ is the gravitational radius ("event horizon") of the Schwarzschild field, $G$ is the gravitational constant, $\rho_K^2 = r^2 + a^2\cos^2\theta$, $\Delta = r^2 - r_0 r + a^2$.

In accordance with the Hilbert causality condition $(g_{00} > 0)$, the metric (6) implies that following inequality holds:

$$\left(1 - \frac{r_0 r}{\rho_K^2}\right) > 0. \qquad (7)$$

In the $(r,\theta)$ coordinates, the equality of (7) to zero determines the outer and the inner surfaces of the ergospheres of the Kerr field.

The contravariant tensor $g^{\alpha\beta}$ has the following form:

[3] The Greek letters take on values 0, 1, 2, 3 and the Roman letters take on values 1, 2, 3.

$$
g^{\alpha\beta} = \left[\begin{array}{c:c:c:c}
\frac{1}{\Delta}\left(r^2 + a^2 + \frac{a^2 r_0 r}{\rho_K^2}\sin^2\theta\right) & 0 & 0 & \frac{a r_0 r}{\Delta\rho_K^2} \\ \hdashline
0 & -\frac{\Delta}{\rho_K^2} & 0 & 0 \\ \hdashline
0 & 0 & -\frac{1}{\rho_K^2} & 0 \\ \hdashline
\frac{a r_0 r}{\Delta\rho_K^2} & 0 & 0 & -\frac{1}{\Delta\sin^2\theta}\left(1 - \frac{r_0 r}{\rho_K^2}\right)
\end{array}\right] \tag{8}
$$

## 3. Hamiltonians of spin-half particles in the Kerr field

In [12], for the metric (6) we obtained a self-adjoint Hamiltonian in the $\eta$ representation with a flat scalar product of the wave functions

$$
\begin{aligned}
H_\eta = {} & \frac{m}{\sqrt{g^{00}}}\gamma^{\underline{0}} - \frac{i\sqrt{\Delta}}{\rho_K\sqrt{g^{00}}}\gamma^{\underline{0}}\gamma^{\underline{1}}\left(\frac{\partial}{\partial r} + \frac{1}{r}\right) - \frac{i}{\rho_K\sqrt{g^{00}}}\gamma^{\underline{0}}\gamma^{\underline{2}}\left(\frac{\partial}{\partial\theta} + \frac{1}{2}\mathrm{ctg}\theta\right) - \\
& - \frac{i}{g^{00}\sqrt{\Delta}\sin\theta}\gamma^{\underline{0}}\gamma^{\underline{3}}\frac{\partial}{\partial\varphi} - \frac{i}{g^{00}}\frac{a r_0 r}{\rho_K^2\Delta}\frac{\partial}{\partial\varphi} - \frac{i}{2}\gamma^{\underline{0}}\gamma^{\underline{1}}\left[\frac{\partial}{\partial r}\frac{\sqrt{\Delta}}{\rho_K\sqrt{g^{00}}}\right] - \\
& - \frac{i}{2}\gamma^{\underline{0}}\gamma^{\underline{2}}\left[\frac{\partial}{\partial\theta}\frac{1}{\rho_K\sqrt{g^{00}}}\right] + \frac{i}{4}\gamma^{\underline{3}}\gamma^{\underline{1}}\sqrt{g^{00}}\frac{\Delta}{\rho_K} a r_0\sin\theta\frac{\partial}{\partial r}\left(\frac{r}{g^{00}\rho_K^2\Delta}\right) - \\
& - \frac{i}{4}\gamma^{\underline{2}}\gamma^{\underline{3}}\sqrt{g^{00}}\frac{\sqrt{\Delta}}{\rho_K} a r_0\sin\theta\frac{\partial}{\partial\theta}\left(\frac{r}{g^{00}\rho_K^2\Delta}\right).
\end{aligned} \tag{9}
$$

If in (9) we restrict ourselves to linear terms with respect to $a$, we will obtain a self-adjoint Hamiltonian for a weak Kerr field

$$
\begin{aligned}
H_\eta^{app} = {} & m\sqrt{f_S}\gamma^{\underline{0}} - i f_S\gamma^{\underline{0}}\gamma^{\underline{1}}\left(\frac{\partial}{\partial r} + \frac{1}{r}\right) - \frac{i r_0}{2r^2}\gamma^{\underline{0}}\gamma^{\underline{1}} - \\
& - i\sqrt{f_S}\gamma^{\underline{0}}\left[\gamma^{\underline{2}}\frac{1}{r}\left(\frac{\partial}{\partial\theta} + \frac{1}{2}\mathrm{ctg}\theta\right) + \gamma^{\underline{3}}\frac{1}{r\sin\theta}\frac{\partial}{\partial\varphi}\right] - \frac{i a r_0}{r^3}\frac{\partial}{\partial\varphi} - \\
& - i\frac{3}{4}\frac{a r_0}{r^3}\sin\theta\gamma^{\underline{3}}\gamma^{\underline{1}}.
\end{aligned} \tag{10}
$$

In (10), $f_S = 1 - \frac{r_0}{r}$. With $a = 0$, the Hamiltonians (9), (10) coincide with the self-adjoint Hamiltonian for the Schwarzschild field [12].

As distinct from centrally symmetric Schwarzschild, Reissner-Nordström and other gravitational fields, the axially symmetric Kerr field does not allow us to use spin-half spherical harmonics for separating angular variables in Dirac’s equation.

Chandrasekhar in his paper [1] separated variables in the Dirac Hamiltonian with the metric (6) using the Penrose-Newman two-component spinor formalism [14] and the tetrad of Kinnersley [15].

Following [16], [17] Dirac's equation and Chadrasekhar's Hamiltonian can be written in the bispinor form:

$$i\frac{\partial\psi_{Ch}}{\partial t}=H_{Ch}\psi_{Ch}=\left(\frac{m}{g^{00}}\gamma^{0}-\frac{i}{g^{00}}\gamma^{0}\gamma^{k}\frac{\partial}{\partial x^{k}}-i\Phi^{0}-\frac{i}{g^{00}}\gamma^{0}\gamma^{k}\Phi^{k}\right)\psi_{Ch}. \tag{11}$$

In (11), $k=1,2,3$, $\Phi^{0}$, $\Phi^{k}$ are bispinor connectivities calculated in a standard way.

The remaining quantities have the following form:

$$\psi_{Ch}=\begin{pmatrix}P^{A}\\ Q_{B}^{*}\end{pmatrix}. \tag{12}$$

In (11),

$$\gamma^{\mu}=\begin{pmatrix}0 & \sqrt{2}\sigma^{\mu AB'}\\ \sqrt{2}\left[\sigma_{AB'}^{\mu}\right]^{T} & 0\end{pmatrix}, \tag{13}$$

$$\sqrt{2}\left[\sigma^{\mu AB'}\right]=\sqrt{2}\begin{pmatrix}n^{\mu} & -m^{*\mu}\\ -m^{\mu} & l^{\mu}\end{pmatrix};\ \sqrt{2}\left[\sigma_{AB'}^{\mu}\right]^{T}=\sqrt{2}\begin{pmatrix}l^{\mu} & m^{*\mu}\\ m^{\mu} & n^{\mu}\end{pmatrix}. \tag{14}$$

In (13), (14), the indices $A,B'$ assume the values of $0$ and $1$, and the symbols $*$ and $T$ mean complex conjugation and transposition.

Components of the tetrad of Kinnersley equal

$$\begin{gathered}l^{\mu}=\frac{1}{\sqrt{2}}\left(\frac{r^{2}+a^{2}}{\Delta},1,0,\frac{a}{\Delta}\right),\ n^{\mu}=\frac{1}{\sqrt{2}\rho_{K}^{2}}\left[r^{2}+a^{2},-\Delta,0,a\right]\\ m^{\mu}=\frac{1}{\sqrt{2}\left(r+ia\cos\theta\right)}\left(ia\sin\theta,0,1,\frac{i}{\sin\theta}\right).\end{gathered} \tag{15}$$

The inverse metric tensor (8) is expressed in terms of the components of (15) as follows:

$$g^{\mu\nu}=l^{\mu}n^{\nu}+n^{\mu}l^{\nu}-m^{\mu}m^{*\nu}-m^{*\mu}m^{\nu}. \tag{16}$$

Chandrasekhar's Hamiltonian (11) is physically equivalent to the self-adjoint Hamiltonian (9), because they are related by a similarity transformation.

For the wave functions of Dirac's equation with Chandrasekhar's Hamiltonian (11), the scalar product contains Parker's weight operator [9], [10] - [12].

$$\rho_{P}=\sqrt{g_{G}}\gamma_{-}^{0}\gamma^{0} \tag{17}$$

where, for the Kerr metric in the Boyer-Lindquist coordinates, $g_{G}=\frac{\rho_{K}^{4}}{r^{4}}$ [12].

If the self-adjoint Hamiltonian (9) is used, $\rho_{P}=1$.

If we define the operator $\eta$ from the equality

$$\rho_P = \eta^+ \eta, \tag{18}$$

Chandrasekhar's Hamiltonian will be related to the Hamiltonian (9) by the similarity transformation

$$H_\eta = \eta H_{Ch} \eta^{-1} = \frac{m}{g^{00}} \gamma_\eta^0 - \frac{i}{g^{00}} \gamma_\eta^0 \gamma_\eta^k \frac{\partial}{\partial x^k} - i\Phi_\eta^0 - \frac{i}{g^{00}} \gamma_\eta^0 \gamma_\eta^k \Phi_\eta^k. \tag{19}$$

In (19)

$$\gamma_\eta^\mu = \eta \gamma^\mu \eta^{-1}, \quad \Phi_\eta^\mu = \eta \Phi^\mu \eta^{-1}. \tag{20}$$

It follows from (19) that both Hamiltonians have the same energy spectrum.

## 4. Physical equivalence of Chandrasekhar's Hamiltonian (11) and the self-adjoint Hamiltonian (9)

We prove the equivalence in the following way.

First, we derive Parker's weight operator from the form of the matrices (13), (14) with the Kinnersley tetrad components (15) and define the transformation operator $\eta$ by means of (18).

We will show that the matrices $\gamma_\eta^0, \gamma_\eta^2$ coincide with the matrices $\tilde{\gamma}^0, \tilde{\gamma}^2$ calculated in [12] by means of tetrads in the Schwinger gauge. For $\gamma_\eta^1, \gamma_\eta^3$ to coincide with $\tilde{\gamma}^1, \tilde{\gamma}^3$, an additional unitary transformation is required, which commutes with the matrices $\gamma^{\underline{0}}, \gamma^{\underline{2}}$ and is related to a spatial turn around the $\theta$ axis.

We denote a part of the Hamiltonian (19) without bispinor connectivities $\Phi_\eta^\mu$ as

$$H_{red} = \frac{m}{g^{00}} \gamma_\eta^0 - \frac{i}{g^{00}} \gamma_\eta^0 \gamma_\eta^k \frac{\partial}{\partial x^k}. \tag{21}$$

The expression (21) is sufficient for us to arrive at the form of the sought self-adjoint Hamiltonian (9) using the formula proven in [12] for the metrics of the form (6) without calculating the bispinor connectivities $\Phi_\eta^\mu$:

$$H_\eta = \frac{1}{2}\left(H_{red} + H_{red}^+\right) + \frac{i}{4}\left(\frac{\partial \tilde{H}_{\underline{n}o}}{\partial x^p} + \frac{g^{ok}}{g^{00}} \frac{\partial \tilde{H}_{\underline{n}k}}{\partial x^p}\right) \tilde{H}_{\underline{m}}^{\,p} S^{\underline{mn}}. \tag{22}$$

In (22), $\tilde{H}_{\underline{n}o}, \tilde{H}_{\underline{n}k}, \tilde{H}_{\underline{m}}^{\,p}$ are tetrads in the Schwinger gauge, $S^{\underline{mn}} = \frac{1}{2}\left(\gamma^{\underline{m}} \gamma^{\underline{n}} - \gamma^{\underline{n}} \gamma^{\underline{m}}\right)$.

As a result, we prove that the two Hamiltonians under consideration are physically equivalent.

Below we consider three cases: the general case of a Kerr field, a weak Kerr field accounting for the terms not higher than linear with respect to the angular momentum parameter $a$, and the Schwarzschild case with $a = 0$.

The matrices (13) with the components of the Kinnersley tetrad (15) have the following form:

$$\gamma^{\mu} = \begin{pmatrix} 0 & \gamma^{\mu}_{up} \\ \gamma^{\mu}_{d} & 0 \end{pmatrix}. \tag{23}$$

In (23), $\gamma^{\mu}_{up}, \gamma^{\mu}_{d}$ are 2 x 2 matrices

$$\gamma^{0}_{up} = \begin{pmatrix} \dfrac{r^2 + a^2}{\rho_K^2} & \dfrac{ia\sin\theta}{\bar{\rho}_K^*} \\ -\dfrac{ia\sin\theta}{\bar{\rho}_K} & \dfrac{r^2 + a^2}{\Delta} \end{pmatrix}, \tag{24}$$

$$\gamma^{0}_{d} = \begin{pmatrix} \dfrac{r^2 + a^2}{\Delta} & -\dfrac{ia\sin\theta}{\bar{\rho}_K^*} \\ \dfrac{ia\sin\theta}{\bar{\rho}_K} & \dfrac{r^2 + a^2}{\rho_K^2} \end{pmatrix}. \tag{25}$$

$$\gamma^{1}_{up} = \begin{pmatrix} -\dfrac{\Delta}{\rho_K^2} & 0 \\ 0 & 1 \end{pmatrix}, \tag{26}$$

$$\gamma^{1}_{d} = \begin{pmatrix} 1 & 0 \\ 0 & -\dfrac{\Delta}{\rho_K^2} \end{pmatrix}, \tag{27}$$

$$\gamma^{2}_{up} = \begin{pmatrix} 0 & -\dfrac{1}{\bar{\rho}_K^*} \\ -\dfrac{1}{\bar{\rho}_K} & 0 \end{pmatrix}, \tag{28}$$

$$\gamma^{2}_{d} = \begin{pmatrix} 0 & \dfrac{1}{\bar{\rho}_K^*} \\ \dfrac{1}{\bar{\rho}_K} & 0 \end{pmatrix}, \tag{29}$$

$$\gamma^{3}_{up} = \begin{pmatrix} \dfrac{a}{\rho_K^2} & \dfrac{i}{\bar{\rho}_K^* \sin\theta} \\ -\dfrac{i}{\bar{\rho}_K \sin\theta} & \dfrac{a}{\Delta} \end{pmatrix}, \tag{30}$$

$$\gamma^{3}_{d} = \begin{pmatrix} \dfrac{a}{\Delta} & -\dfrac{i}{\bar{\rho}_K^* \sin\theta} \\ \dfrac{i}{\bar{\rho}_K \sin\theta} & \dfrac{a}{\rho_K^2} \end{pmatrix}. \tag{31}$$

In (24) - (31), $\bar{\rho}_K = r + ia\cos\theta$; $\bar{\rho}_K^* = r - ia\cos\theta$.

The expressions for $\gamma^{\mu}$ for the weak Kerr field and the Schwarzschild field $(a=0)$ are derived from (24) - (31) with $\Delta \to r^2 f_S = r^2\left(1-\frac{r_0}{r}\right)$; $\rho_K^2 \to r^2$.

With transition to the flat Minkowski space $(r_0 = 0, a = 0)$, the matrices (24) - (31) correspond to Dirac's $\gamma$-matrices in the spinor representation

$$\gamma^{\underline{0}} = \begin{pmatrix} 0 & 1 \\ 1 & 0 \end{pmatrix}; \gamma^{\underline{i}} = \begin{pmatrix} 0 & -\sigma^i \\ \sigma^i & 0 \end{pmatrix}. \tag{32}$$

Parker's weight operator is

$$\rho_P = \frac{\rho_K^2}{r^2}\gamma^{\underline{0}}\gamma^0 = \frac{\rho_K^2}{r^2}\begin{pmatrix} \rho_P^{up} & 0 \\ 0 & \rho_P^d \end{pmatrix}, \tag{33}$$

where $\rho_P^{up}, \rho_P^d$ are 2 x 2 matrices.

$$\rho_P^{up} = \begin{pmatrix} \dfrac{r^2+a^2}{\Delta} & -\dfrac{ia\sin\theta}{\bar{\rho}_K^*} \\ \dfrac{ia\sin\theta}{\bar{\rho}_K} & \dfrac{r^2+a^2}{\rho_K^2} \end{pmatrix}, \tag{34}$$

$$\rho_P^d = \begin{pmatrix} \dfrac{r^2+a^2}{\rho_K^2} & \dfrac{ia\sin\theta}{\bar{\rho}_K^*} \\ -\dfrac{ia\sin\theta}{\bar{\rho}_K} & \dfrac{r^2+a^2}{\Delta} \end{pmatrix}. \tag{35}$$

The expressions for $\rho_P^{up}, \rho_P^d$ for the weak Kerr field and the Schwarzschild field are easily established from (34), (35).

The operator of transformation to the $\eta$ representation must satisfy the equality

$$\rho_P = \eta^+\eta, \tag{36}$$

where

$$\eta = \frac{\sqrt{\rho_K^2}}{r}\begin{pmatrix} \eta_{up} & 0 \\ 0 & \eta_d \end{pmatrix}, \tag{37}$$

$\eta_{up}, \eta_d$ are 2 x 2 matrices.

The operator $\eta$ can be defined by reducing the matrix $\rho_P$ (33) to the diagonal form. In this case the eigenvalues and the eigenvectors of this 4x4 matrix have to be determined.

However, taking into account the block-diagonal structure (33), this task for $\rho_p$ is reduced to the calculating of the eigenvalues and the eigenvectors for 2x2 matrices $\rho_P^{up}$ (34), $\rho_P^d$ (35).

For the matrices (34), (35) the eigenvalues are identical and have the following form

$$\lambda_1 = \frac{\left(r^2+a^2\right)\left(\Delta+\rho_K^2\right)-B}{2\Delta\rho_K^2}, \quad (38)$$

$$\lambda_2 = \frac{\left(r^2+a^2\right)\left(\Delta+\rho_K^2\right)+B}{2\Delta\rho_K^2}, \quad (39)$$

where $B=\sqrt{A^2+4\Delta^2\rho_K^2 a^2\sin^2\theta}, \quad A=\left(r^2+a^2\right)\left(\rho_K^2-\Delta\right)$.

Note that

$$\lambda_1\lambda_2 = g^{00}. \quad (40)$$

The unnormalized eigenvectors $\mathbf{U},\mathbf{V}$ of the matrices $\rho_P^{up}$ (34), $\rho_P^d$ (35) are identical and can be represent in the form

$$\mathbf{U}=\begin{pmatrix} U_1 \\ U_2 \end{pmatrix}=\begin{pmatrix} \dfrac{2ia\Delta\bar{\rho}_K\sin\theta}{A+B} \\ 1 \end{pmatrix}, \quad \mathbf{V}=\begin{pmatrix} V_1 \\ V_2 \end{pmatrix}=\begin{pmatrix} -\dfrac{2ia\Delta\bar{\rho}_K\sin\theta}{B-A} \\ 1 \end{pmatrix}. \quad (41)$$

Consequently, the following equalities are fulfilled

$$\rho_p^{up}\begin{pmatrix} U_1 \\ U_2 \end{pmatrix}=\lambda_1\begin{pmatrix} U_1 \\ U_2 \end{pmatrix}, \quad \rho_p^{up}\begin{pmatrix} V_1 \\ V_2 \end{pmatrix}=\lambda_2\begin{pmatrix} V_1 \\ V_2 \end{pmatrix}. \quad (42)$$

$$\rho_p^{d}\begin{pmatrix} U_1 \\ U_2 \end{pmatrix}=\lambda_2\begin{pmatrix} U_1 \\ U_2 \end{pmatrix}, \quad \rho_p^{d}\begin{pmatrix} V_1 \\ V_2 \end{pmatrix}=\lambda_1\begin{pmatrix} V_1 \\ V_2 \end{pmatrix}. \quad (43)$$

In terms of 4x4 matrix $\rho_p$ these relations are equivalent the following ones

$$\rho_p\begin{pmatrix} U_1 \\ U_2 \\ 0 \\ 0 \end{pmatrix}=\frac{\rho_k^2}{r^2}\lambda_1\begin{pmatrix} U_1 \\ U_2 \\ 0 \\ 0 \end{pmatrix}; \quad \rho_p\begin{pmatrix} V_1 \\ V_2 \\ 0 \\ 0 \end{pmatrix}=\frac{\rho_k^2}{r^2}\lambda_2\begin{pmatrix} V_1 \\ V_2 \\ 0 \\ 0 \end{pmatrix}; \quad (44)$$

$$\rho_p\begin{pmatrix} 0 \\ 0 \\ U_1 \\ U_2 \end{pmatrix}=\frac{\rho_k^2}{r^2}\lambda_2\begin{pmatrix} 0 \\ 0 \\ U_1 \\ U_2 \end{pmatrix}; \quad \rho_p\begin{pmatrix} 0 \\ 0 \\ V_1 \\ V_2 \end{pmatrix}=\frac{\rho_k^2}{r^2}\lambda_1\begin{pmatrix} 0 \\ 0 \\ V_1 \\ V_2 \end{pmatrix}. \quad (45)$$

The operators $\eta_{up}^+, \eta_{up}, \eta_d^+, \eta_d$ then equal

$$\eta_{up}^{+} = \begin{pmatrix} N_2 V_1 & N_1 U_1 \\ N_2 V_2 & N_1 U_2 \end{pmatrix} \begin{pmatrix} \sqrt{\lambda_2} & 0 \\ 0 & \sqrt{\lambda_1} \end{pmatrix}, \tag{46}$$

$$\eta_{up} = \begin{pmatrix} \sqrt{\lambda_2} & 0 \\ 0 & \sqrt{\lambda_1} \end{pmatrix} \begin{pmatrix} N_2^* V_1^* & N_2^* V_2^* \\ N_1^* U_1^* & N_1^* U_2^* \end{pmatrix}, \tag{47}$$

$$\eta_{d}^{+} = \begin{pmatrix} N_2 V_1 & N_1 U_1 \\ N_2 V_2 & N_1 U_2 \end{pmatrix} \begin{pmatrix} \sqrt{\lambda_1} & 0 \\ 0 & \sqrt{\lambda_2} \end{pmatrix}, \tag{48}$$

$$\eta_{d} = \begin{pmatrix} \sqrt{\lambda_1} & 0 \\ 0 & \sqrt{\lambda_2} \end{pmatrix} \begin{pmatrix} N_2^* V_1^* & N_2^* V_2^* \\ N_1^* U_1^* & N_1^* U_2^* \end{pmatrix}. \tag{49}$$

The normalization coefficients $N_1, N_2$ in (46) - (49) are determined from the condition (36):

$$N_1 = \sqrt{\frac{B+A}{2B}};\ N_2 = -i\sqrt{\frac{B-A}{2B}}. \tag{50}$$

The phase factors in $N_1, N_2$ are chosen subject to correct limit transition to the operator $\eta$ for the Schwarzschild field with angular momentum parameter $a = 0$.

Considering (41), the operators $\eta_{up}, \eta_d$ in (37) and $\eta_{up}^{-1}, \eta_d^{-1}$ can be written in the form

$$\eta_{up} = \begin{pmatrix} N_1 \frac{\bar{\rho}_K^*}{\sqrt{\rho_K^2}} \sqrt{\lambda_2} & N_2 \sqrt{\lambda_2} \\ N_2 \frac{\bar{\rho}_K^*}{\sqrt{\rho_K^2}} \sqrt{\lambda_1} & N_1 \sqrt{\lambda_1} \end{pmatrix}, \tag{51}$$

$$\eta_{up}^{-1} = \begin{pmatrix} N_1 \frac{\bar{\rho}_K}{\sqrt{\rho_K^2}} \frac{1}{\sqrt{\lambda_2}} & N_2^* \frac{\bar{\rho}_K}{\sqrt{\rho_K^2}} \frac{1}{\sqrt{\lambda_1}} \\ N_2^* \frac{1}{\sqrt{\lambda_2}} & N_1 \frac{1}{\sqrt{\lambda_1}} \end{pmatrix}. \tag{52}$$

$$\eta_{d} = \begin{pmatrix} N_1 \frac{\bar{\rho}_K^*}{\sqrt{\rho_K^2}} \sqrt{\lambda_1} & N_2 \sqrt{\lambda_1} \\ N_2 \frac{\bar{\rho}_K^*}{\sqrt{\rho_K^2}} \sqrt{\lambda_2} & N_1 \sqrt{\lambda_2} \end{pmatrix}, \tag{53}$$

$$\eta_{d}^{-1} = \begin{pmatrix} N_1 \frac{\bar{\rho}_K}{\sqrt{\rho_K^2}} \frac{1}{\sqrt{\lambda_1}} & N_2^* \frac{\bar{\rho}_K}{\sqrt{\rho_K^2}} \frac{1}{\sqrt{\lambda_2}} \\ N_2^* \frac{1}{\sqrt{\lambda_1}} & N_1 \frac{1}{\sqrt{\lambda_2}} \end{pmatrix}. \tag{54}$$

For the weak Kerr field,

$$\lambda_1 = 1; \lambda_2 = \frac{1}{f_S} = \frac{1}{1-\frac{r_0}{r}}; N_1 = 1; N_2 = -i\frac{\sqrt{f_S}\, a\sin\theta}{r(1-f_S)};$$

$$\eta_{up} = \begin{pmatrix} \frac{1}{\sqrt{f_S}} & -i\frac{\sqrt{f_S}\, a\sin\theta}{r(1-f_S)} \\ -i\frac{f_S a\sin\theta}{r(1-f_S)} & 1 \end{pmatrix}; \eta_{up}^{-1} = \begin{pmatrix} \sqrt{f_S} & i\frac{f_S a\sin\theta}{r(1-f_S)} \\ i\frac{f_S^{3/2} a\sin\theta}{r(1-f_S)} & 1 \end{pmatrix}, \tag{55}$$

$$\eta_d = \begin{pmatrix} 1 & -i\frac{f_S a\sin\theta}{r(1-f_S)} \\ -i\frac{\sqrt{f_S}\, a\sin\theta}{r(1-f_S)} & \frac{1}{\sqrt{f_S}} \end{pmatrix}; \eta_d^{-1} = \begin{pmatrix} 1 & i\frac{f_S^{3/2} a\sin\theta}{r(1-f_S)} \\ i\frac{f_S a\sin\theta}{r(1-f_S)} & \sqrt{f_S} \end{pmatrix}. \tag{56}$$

The transformation matrix $\eta$ for the Schwarzschild field $(a=0)$ is diagonal,

$$\eta_S = diag\left[\frac{1}{\sqrt{f_S}}, 1, 1, \frac{1}{\sqrt{f_S}}\right]. \tag{57}$$

Next, we transform the matrices (24) - (31). Following the transformations, we perform an equivalent cyclic replacement of Dirac's local matrices $\underline{\gamma}^3 \to \underline{\gamma}^1; \underline{\gamma}^1 \to \underline{\gamma}^2; \underline{\gamma}^2 \to \underline{\gamma}^3$. As a result, we have

$$\eta\gamma^0\eta^{-1} = \sqrt{g^{00}}\,\underline{\gamma}^0, \tag{58}$$

$$\eta\gamma^1\eta^{-1} = \frac{\sqrt{\Delta}}{\sqrt{\rho_K^2}}\left(\underline{\gamma}^1\cos\delta - \underline{\gamma}^3\sin\delta\right), \tag{59}$$

$$\eta\gamma^2\eta^{-1} = \frac{1}{\sqrt{\rho_K^2}}\underline{\gamma}^2, \tag{60}$$

$$\eta\gamma^3\eta^{-1} = \frac{1}{\sin\theta\sqrt{\Delta}\sqrt{g^{00}}}\left(\underline{\gamma}^3\cos\delta + \underline{\gamma}^1\sin\delta\right) + \frac{a r_0 r}{\rho_K^2\Delta\left(g^{00}\right)^{3/2}}\underline{\gamma}^0. \tag{61}$$

In (59), (61)

$$\cos\delta = -\frac{\sqrt{g^{00}}\sqrt{\Delta}\sqrt{\rho_K^2}\left(\Delta - \rho_K^2\right)}{B}, \tag{62}$$

$$\sin\delta = \frac{a\sin\theta\sqrt{\Delta}\left(\Delta + \rho_K^2\right)}{B}. \tag{63}$$

For the weak Kerr field,

$$\cos\delta = 1; \quad \sin\delta = \frac{\sqrt{f_S}}{r}\frac{(1+f_S)}{(1-f_S)}a\sin\theta. \tag{64}$$

For the Schwarzschild field

$$\cos\delta = 1; \quad \sin\delta = 0. \tag{65}$$

The transformed matrices (58), (60) coincide with the matrices derived in [12] using tetrads in the Schwinger gauge. The matrices (59), (61) differ from the similar matrices in [12] in the spatial turn around the $\theta$ axis. The coincidence of the matrices can be provided by applying an additional unitary transformation

$$R = e^{\frac{\delta}{2}\gamma^{\underline{3}}\gamma^{\underline{1}}} = \cos\frac{\delta}{2} - i\Sigma^{\underline{2}}\sin\frac{\delta}{2}. \tag{66}$$

The transformation (66) commutes with the matrices $\gamma^{\underline{0}},\gamma^{\underline{2}}$; $-i\Sigma^{\underline{2}} = \gamma^{\underline{3}}\gamma^{\underline{1}}$.

Finally we have

$$R\eta\gamma^{0}\eta^{-1}R^{-1} = \sqrt{g^{00}}\,\gamma^{\underline{0}}, \tag{67}$$

$$R\eta\gamma^{1}\eta^{-1}R^{-1} = \frac{\sqrt{\Delta}}{\sqrt{\rho_K^2}}\gamma^{\underline{1}}, \tag{68}$$

$$R\eta\gamma^{2}\eta^{-1}R^{-1} = \frac{1}{\sqrt{\rho_K^2}}\gamma^{\underline{2}}, \tag{69}$$

$$R\eta\gamma^{3}\eta^{-1}R^{-1} = \frac{1}{\sin\theta\sqrt{\Delta}\sqrt{\rho_K^2}}\gamma^{\underline{3}} + \frac{a r_0 r}{\rho_K^2\Delta\left(g^{00}\right)^{3/2}}\gamma^{\underline{0}}. \tag{70}$$

The equalities (67) - (70) together with the expressions (21), (22) prove that Chandrasekhar's Hamiltonian (11) and the self-adjoint Hamiltonian are physically equivalent.

## 5. Hermiticity of Dirac Hamiltonians in the Kerr field

The standard condition of Hermiticity of Dirac Hamiltonians in external gravitational fields expressed in terms of scalar products of the wave functions has the form

$$(\psi, H\varphi) = (H\psi, \varphi). \tag{71}$$

In [10], the condition (71) is transformed into

$$\oint dS_k\left(\sqrt{-g}\, j^k\right) + \int d^3x\sqrt{-g}\left[\psi^{+}\gamma^{\underline{0}}\left(\gamma^{0}_{,0} + \left(\begin{matrix} 0 \\ 0\ 0 \end{matrix}\right)\gamma^{0}\right)\psi + \left(\begin{matrix} k \\ k\ 0 \end{matrix}\right) j^{0}\right] = 0. \tag{72}$$

The equality (72) implies summation over $k$, where $k = 1,2,3$; $\oint dS_k$ is the integral over projections of the surfaces bounding the domain of the wave functions $\psi,\psi^{+}$ of Dirac's equation; $g$ is the determinant of the corresponding metric, for the Kerr metric $(-g) = \rho_K^4\sin^2\theta$; $\gamma^{0}_{,0} \equiv \frac{\partial\gamma^{0}}{\partial x^{0}}$; $\left(\begin{matrix} 0 \\ 0\ 0 \end{matrix}\right), \left(\begin{matrix} k \\ k\ 0 \end{matrix}\right)$ are Christoffel symbols; $j^{\mu}$ is a four-dimensional vector equal to

$$j^{\mu} = \psi^{+}\gamma^{\underline{0}}\gamma^{\mu}\psi. \tag{73}$$

For the stationary Kerr field, $\gamma^0_{,0}=0$; the Christoffel symbols $\begin{pmatrix} 0 \\ 0\;0 \end{pmatrix}, \begin{pmatrix} k \\ k\;0 \end{pmatrix}=0$, and the condition (72) becomes equal to

$$\oint dS_k\left(\sqrt{-g}\,j^k\right)=0. \tag{74}$$

For the stationary case, the wave function in Dirac's equation (11) with Chandrasekhar's Hamiltonian can be written as

$$\psi_{Ch}(\mathbf{r},t)=\psi_{Ch}(\mathbf{r})e^{-iEt}, \tag{75}$$

where $E$ is the energy of a Dirac particle.

Chandrasekhar in [1] represented (75) in the form

$$\psi_{Ch}(\mathbf{r},t)=\begin{pmatrix} \dfrac{1}{r-ia\cos\theta}\overset{(-)}{R}(r)\overset{(-)}{S}(\theta) \\ \dfrac{1}{\sqrt{\Delta}}\overset{(+)}{R}(r)\overset{(+)}{S}(\theta) \\ -\dfrac{1}{\sqrt{\Delta}}\overset{(+)}{R}(r)\overset{(-)}{S}(\theta) \\ -\dfrac{1}{r+ia\cos\theta}\overset{(-)}{R}(r)\overset{(+)}{S}(\theta) \end{pmatrix} e^{-iEt}e^{im_\varphi\varphi} \tag{76}$$

and obtained separately two equations for the angular functions $\overset{(-)}{S}(\theta), \overset{(+)}{S}(\theta)$ and two equations for the radial functions $\overset{(-)}{R}(r), \overset{(+)}{R}(r)$. In (76), $m_\varphi$ is the magnetic quantum number.

Using the function (76), for each value of $E, m_\varphi, a$ and for the quantum numbers of a particle's orbital and total momentum $l, j$, we can define current components $j^\mu$ using the matrices $\gamma^\mu$ (23) - (31). As a result, we have

$$\begin{aligned} j^0(r,\theta)&=\frac{2\left(r^2+a^2\right)}{\rho_K^2\Delta}\overset{(+)}{R}_{lm_\varphi}(r)\overset{(-)}{R}_{lm_\varphi}(r)\left(\overset{(+)}{S}_{lm_\varphi}(\theta)\overset{(+)}{S}_{lm_\varphi}(\theta)+\overset{(-)}{S}_{lm_\varphi}(\theta)\overset{(-)}{S}_{lm_\varphi}(\theta)\right)+ \\ &+2\frac{ia\sin\theta}{\rho_K^2\sqrt{\Delta}}\left(\overset{(-)}{R}_{lm_\varphi}(r)\overset{(-)}{R}_{lm_\varphi}(r)-\overset{(+)}{R}_{lm_\varphi}(r)\overset{(+)}{R}_{lm_\varphi}(r)\right)\overset{(+)}{S}_{lm_\varphi}(\theta)\overset{(-)}{S}_{lm_\varphi}(\theta)= \\ &=\frac{r^2+a^2}{2\rho_K^2\Delta}\left(g^2_{lm_\varphi}(r)+f^2_{lm_\varphi}(r)\right)\left(\overset{(+)}{S}_{lm_\varphi}(\theta)\overset{(+)}{S}_{lm_\varphi}(\theta)+\overset{(-)}{S}_{lm_\varphi}(\theta)\overset{(-)}{S}_{lm_\varphi}(\theta)\right)- \\ &-\frac{2a\sin\theta}{\rho_K^2\sqrt{\Delta}}g_{lm_\varphi}(r)f_{lm_\varphi}(r)\overset{(+)}{S}_{lm_\varphi}(\theta)\overset{(-)}{S}_{lm_\varphi}(\theta), \end{aligned} \tag{77}$$

$$j^r(r,\theta)=0, \tag{78}$$

$$j^\theta(r,\theta)=0, \tag{79}$$

$$j^{\varphi}(r,\theta) = i\frac{2}{\rho_K^2\sqrt{\Delta}\sin\theta}\left(\overset{(-)}{R}_{lm_\varphi}(r)\overset{(-)}{R}_{lm_\varphi}(r) - \overset{(+)}{R}_{lm_\varphi}(r)\overset{(+)}{R}_{lm_\varphi}(r)\right)\overset{(+)}{S}_{lm_\varphi}(\theta)\overset{(-)}{S}_{lm_\varphi}(\theta) +$$
$$+\frac{2a}{\rho_K^2\Delta}\overset{(+)}{R}_{lm_\varphi}(r)\overset{(-)}{R}_{lm_\varphi}(r)\left(\overset{(+)}{S}_{lm_\varphi}(\theta)\overset{(+)}{S}_{lm_\varphi}(\theta) + \overset{(-)}{S}_{lm_\varphi}(\theta)\overset{(-)}{S}_{lm_\varphi}(\theta)\right) =$$
$$= -\frac{2}{\rho_K^2\sqrt{\Delta}\sin\theta}g_{lm_\varphi}(r)\,f_{lm_\varphi}(r)\overset{(+)}{S}_{lm_\varphi}(\theta)\overset{(-)}{S}_{lm_\varphi}(\theta) + \tag{80}$$
$$+\frac{a}{2\rho_K^2\Delta}\left(g_{lm_\varphi}^2(r) + f_{lm_\varphi}^2(r)\right)\left(\overset{(+)}{S}_{lm_\varphi}(\theta)\overset{(+)}{S}_{lm_\varphi}(\theta) + \overset{(-)}{S}_{lm_\varphi}(\theta)\overset{(-)}{S}_{lm_\varphi}(\theta)\right).$$

When deriving (77) - (80) we used the equality

$$\overset{(-)}{R}{}^{*}_{lm_\varphi}(r) = \overset{(+)}{R}_{lm_\varphi}(r) \tag{81}$$

and reality conditions for the angular functions $\overset{(+)}{S}_{lm_\varphi}, \overset{(-)}{S}_{lm_\varphi}$

$$\begin{gathered}\overset{(+)}{S}{}^{*}_{lm_\varphi} = \overset{(+)}{S}_{lm_\varphi},\\ \overset{(-)}{S}{}^{*}_{lm_\varphi} = \overset{(-)}{S}_{lm_\varphi}.\end{gathered} \tag{82}$$

In (77) - (80), $g(r)\,f(r)$ are real functions:

$$\begin{gathered}g(r) = \overset{(-)}{R}(r) + \overset{(+)}{R}(r),\\ f(r) = -i\left(\overset{(-)}{R}(r) - \overset{(+)}{R}(r)\right).\end{gathered} \tag{83}$$

Considering that the radial (78) and the polar (79) current components $j^{\mu}(r,\theta)$ are zero, the Hermiticity condition (74) becomes equal to

$$\oint dS_{\varphi}\left(\sqrt{-g}\,j^{\varphi}\right) = 0. \tag{84}$$

It follows that for any surface of revolution around the $z$ axis ($F(r,\theta) = 0$) bounding the domain of the wave functions of Dirac's equation, the Hermiticity conditions for Dirac's Hamiltonians in the Kerr field (84) and, consequently, (71), are satisfied.

This also holds for the surfaces of the outer and the inner ergospheres of the Kerr field.

## Conclusions

In the paper we prove that Chandrasekhar's Dirac Hamiltonian (11) and the self-adjoint Hamiltonian (9) with a flat scalar product of the wave functions are physically equivalent.

An explicit form of the operators $\eta, \eta^{-1}$ that transform Chandrasekhar's Hamiltonian and wave functions to the $\eta$ representation is obtained:

$$\begin{aligned} &\psi_\eta(\mathbf{r},t) = \eta\psi_{Ch}(\mathbf{r},t), \\ &H_\eta = \eta H_{Ch}\eta^{-1}. \end{aligned} \tag{85}$$

We show that if the domain of the wave functions of Dirac's equation in the Kerr field is bounded by a two-dimensional surfaces of revolution around the $z$ axis, Chandrasekhar's Hamiltonian (11) and the self-adjoint Hamiltonian (9) are Hermitian with equality of the scalar products,

$$(\psi, H\varphi) = (H\psi, \varphi). \tag{86}$$

In a similar manner one can demonstrate the equivalence and the Hermiticity of Page's Hamiltonian [3] and a corresponding self-adjoint Hamiltonian in the $\eta$ representation in the Kerr-Newman field [4].

## References

[1] S.Chandrasekhar. Proc. Roy. Soc (London), A349, 571, 1976.

[2] R.P.Kerr. Phys. Rev. Lett. 11, 237 (1963).

[3] D.Page. Phys. Rev. D14, 1509, 1976.

[4] E.T.Newman, E.Couch, K.Chinnapared, A.Exton, A.Prakash and R.Torrence. J.Math.Phys.6, 918 (1965).

[5] R.H.Boyer and R.W.Lindquist. J. Math. Phys. 8, 265-281 (1967).

[6] C.M.Bender, D.Brody and H.F.Jones, Phys. Rev. Lett. 89, 270401 (2002); Phys. Rev. D. 70, 025001 (2004).

[7] A.Mostafazadeh, J. Math. Phys. (N.Y.) 43, 205 (2002); 43, 2814 (2002); 43, 3944 (2002); arxiv: 0810.5643v3.

[8] B.Bagchi and A.Fring, Phys. Lett. A 373, 4307 (2009).

[9] L.Parker. Phys. Rev. D22, 1922 (1980).

[10] M.V.Gorbatenko, V.P.Neznamov. Phys. Rev. D 82, 104056 (2010); arxiv: 1007.4631v1 (gr-qc).

[11] M.V.Gorbatenko, V.P.Neznamov. Phys. Rev. D 83, 105002 (2011); arxiv: 1102.4067v1 (gr-qc).

[12] M.V.Gorbatenko, V.P.Neznamov. arxiv: 1107.0844 (gr-qc).

[13] M.V.Gorbatenko, V.P.Neznamov. Ann. Phys. (Berlin), 1–6 (2014)/DOI 10.1002/andp.201300218.

[14] E.T.Newman, R.Penrose. J.Math.Phys. 3, 566 (1962).

[15] W.Kinnersley. J.Math.Phys. 10, 1195 (1969).

[16] I.M.Ternov, A.B.Gaina, G.A.Chizhov. Sov. Phys. J. 23, 695-700, 1980.

[17] S.R.Dolan. Trinity Hall and Astrophysics Group, Cavendish Laboratory. Dissertation, 2006.